\documentclass[preprints,article,accept,pdftex,moreauthors]{Definitions/mdpi} 
\firstpage{1} 
\pubvolume{1}
\issuenum{1}
\articlenumber{0}
\pubyear{2026}
\copyrightyear{2026}
\datereceived{ } 
\daterevised{ } 
\dateaccepted{ } 
\datepublished{ } 

\Title{Ionization potential depression in degenerate plasmas and Pauli blocking of multi-electron ions}

\Author{Gerd R\"opke \orcidA{0000-0001-9319-5959}*}

\AuthorNames{Gerd R\"opke
}

\address{%
Institute of Physics, University of Rostock, Albert-Einstein Str. 23-24, D-18059 Rostock, Germany
}

\corres{Correspondence: gerd.roepke@uni-rostock.de
}

\abstract{The composition of partially ionised plasmas is investigated for densities and temperatures at which the free electrons are degenerate. Based on a quantum statistical approach,
the effect of Pauli blocking is addressed. Specifically, one- and two-electron ions are studied.
Approximations for deriving an in-medium Schr\"odinger equation for the ionization potential are indicated.
New results regarding the degree of ionisation and the Mott effect are presented.
Standard codes for plasma properties do not take Pauli blocking effects into account and are therefore unable to explain the experiments in the high-density regime, where the electrons are degenerate.
}


\begin{document}




\section{Introduction}

The properties of atoms and ions immersed in a dense plasma are altered by medium effects.
An important example is Debye screening, which modifies the energies of charged particles. 
Consequently, the ionisation potential changes in a dense plasma \cite{Unsoeld48}. Simple approaches 
such as the Debye shift, which are valid at low densities, can be extended to higher densities, 
where other approaches such as the ion sphere model are more suitable. Interpolation formulas 
\cite{EK63,SP66}
have been proposed and compared with recent experiments on high-density plasmas 
\cite{Hoarty13,VCW14,Vinko12,Vinko15,ciricosta12,ciricosta16}.
A more fundamental approach \cite{Crowley14}, based on a quantum-statistical many-particle theory 
\cite{KKER,KSK}, is required for a systematic treatment of these plasma effects.
In particular, the dynamic screening by free electrons and the inclusion of the ion structure factor 
were investigated in \cite{Lin17} and further references given there. More recently, Monte Carlo simulations using path integrals \cite{Militzer,Bonitz25,Bellenbaum25} have been employed.
See also the most recent publications on ionization potential depression and ionization balance in dense plasmas \cite{roadmap25,Zeng25,Wu25} for further references.

New experiments with warm, dense matter (WDM) achieve high densities at which the plasma is degenerate, 
see Refs. \cite{preston13,Fletcher14,Kraus16,WDM}. 
For a plasma with temperature $T$ and free electron density $n_e$,
the electron degeneracy parameter is defined as 
\begin {equation}
\label{Theta}
\Theta = \frac{T}{T_{\rm Fermi}}=\frac{2 m_e k_BT}{\hbar^2} (3 \pi^2 n_e)^{-2/3}
\end{equation}
and describes the region $\Theta \le 1$ in the temperature-density plane in which a classical description is no longer valid. Instead, quantum effects, in particular the Pauli principle as a consequence of the antisymmetry of the many-electron wave function, lead to so-called exchange terms. The associated condition
$n_e \Lambda_e^3/2 \ge 1$ (spin factor 2) with 
$\Lambda_e = (2 \pi \hbar^2/m_e k_BT)^{1/2}$
is known as the condition at which the classical Boltzmann gas approximation is no longer applicable and the quantum description based on the Fermi distribution function must be applied. For example: For $T= 100$ eV, the value $\Theta=1$ occurs at a density of $n_e=4.54 \times 10^{24}$ cm$^{-3}$.
At the National Ignition Facility (NIF) and other facilities, new experiments are being conducted and planned to investigate WDM at free electron densities of up to  $10^{25}$ cm$^{-3}$ and  temperatures of up to $k_BT=10^2$ eV, so that the effects of degeneracy are relevant. 
In this work, we restrict  ourselves to non-relativistic plasmas. Relativistic effects become non-negligible at $k_BT  \gtrsim 10^5$ eV and $n_e  \gtrsim 10^{29}$ cm$^{-3}$.
Relativistic plasmas occur in astrophysical objects such as the crust of proto-neutron stars or during neutron star mergers; see Refs. \cite{Barba,Harutyunan}  and further references cited there.

A significant issue is the equation of state under such extreme conditions, particularly the composition. The degree of ionisation increases with rising $T$, but at high densities, pressure ionisation also occurs due to screening effects. Tables that take these effects into account are available, for example OPAL \cite{OPAL,OPAL1,OPAL2,OPAL3}. 
They are based on a many-particle treatment of screening \cite{KKER,KSK,R13}.

In degenerate systems, new physical phenomena become increasingly significant. Whilst at lower densities, in the classical regime, dynamic screening is the most important medium effect, exchange interactions become increasingly important at these extremely high densities.
These Pauli blocking effects are being investigated for Hydrogen plasmas; see \cite{EBRRR09}.
They have been discussed for nuclear matter \cite{R} to describe the dissolution of binding states in nuclear systems at high densities.
A recent discussion of Pauli blocking effects in connection with the $K$-edge shift can be found in \cite{Hu,Rosmej}.
In particular, the ionisation potential depression (IPD) and Pauli blocking in degenerate Carbon plasmas at extreme densities were investigated in the article \cite{IPD19}. Due to Pauli blocking, the degree of ionisation at high densities is greater than predicted by standard approaches such as OPAL.
Similarly, a recent publication \cite{Nature23} found that the ion charge inferred from the scattering data for Beryllium at high densities is significantly higher than widely used  analytical models predict.
We focus on  Carbon; the calculations of Ref. \cite{IPD19} are presented in explicit form in App. \ref{sec:Carbon}, where all details are provided in that section so that the calculations can be reproduced immediately without additional sources. Our aim is to investigate the approximations used in these calculations.
We mention the discrepancy with density functional theory–molecular dynamics (DFT-MD) simulations \cite{Bethkenhagen20},
which are also discussed in this work.

The influence of Pauli blocking on single-electron ions, i.e. H-like bound states, has already been discussed in several papers; 
see \cite{IPD19,EBRRR09,Ebeling23,Ebeling24,Ebeling26} and further references cited therein.
In this work, we focus on multi-electron ions, so that other materials and ions can also be addressed. 
In particular, we discuss the problem of He and He-like bound states.
Our aim is to contribute to the study of hot and dense matter consisting of multi-electron components in the region where the electrons are degenerate.
We provide a brief overview of the in-medium Schrödinger equation (Sect. \ref{sec:2}), the composition of the partially ionised plasma (Sect. \ref{sec:3}) and the treatment of single-electron ions (Sect. \ref{sec:4}). 
A new topic is the treatment of two-electron ions (He-like ions) in Sect. \ref{sec:5}.
A discussion of multi-electron ions, as well as a comparison with other approaches, can be found in Sect. \ref{sec:6}, before we conclude with some final remarks (Sect. \ref{sec:7}).

In this work, we use Rydberg units, where $\hbar=2 m_e=e^2/8 \pi \epsilon_0=1$, so that energies are expressed in Ry, where 1 Ha = 2 Ry = 27.2114 eV = 219474.63 cm$^{-1}$.
The Bohr radius is $a_{\rm B}=5.291772 \times 10^{-9}$ cm.
We use atomic units $T_{\rm Ry}$ for the temperature $T$ and $n_{\rm Bohr} = n\,a^3_{\rm B}$ for the particle density, 
so that
\begin{equation}
\label{units}
    T_{\rm Ry}=6.33358 \times 10^{-6}\, \frac{T}{{\rm K}}, \quad n_{\rm Bohr}= 1.4818471\times 10^{-25} n_{\rm e}\,{\rm cm}^3\,.
\end{equation}
The mass density includes the mass of the nucleus; for $^{12}$C, $\rho_{\rm C}= 134.481 \times n_{\rm nucl} a_B^3$ g/cm$^3$.

\section{Multi-electron Schr\"odinger Equation}
\label{sec:2}

\subsection{Free multi-electron Schr\"odinger Equation}

We consider warm and dense matter composed of elements with proton number $Z$, for example $Z=1$ (Hydrogen), $Z=2$ (Helium), or $Z=6$ (Carbon). 
For neutral matter, the total density of nuclei $n_{\rm nucl}$ and the total number density of electrons $n^{\rm tot}_e$ are related as follows: $n^{\rm tot}_e=Z\,n_{\rm nucl}$. 
The interaction is described by Coulomb’s law $Q_1Q_2/(4 \pi \epsilon_0 r)$, in the momentum representation
 \begin{equation}
V_{Q_1,Q_2}(k_1,k_2;k'_1,k'_2)= \frac{Q_1Q_2}{\Omega_0\epsilon_0 |{\bf k}_1-{\bf k}_1'|^2}\delta_{{\bf k}_1+{\bf k}_2,{\bf k}'_1+{\bf k}'_2} \delta_{\sigma_1,\sigma_1'} \delta_{\sigma_2,\sigma_2'}     
 \end{equation}
where $Q_i=Z\,e$ applies to nuclei with mass $m_Z$ and $Q_i=-e$ applies to electrons with mass $m_e$; $\Omega_0$ denotes the system volume. The quantum numbers $k={\bf k},\sigma$ represent the wave number vector and the spin orientation of the interacting particles.

This interaction can give rise to bound states, which are determined by solving the Schr\"odinger equation.
For example, for the isolated electron-proton system ($Z=0$), the energy levels are $E^0_n=-1/n^2$ Ry, where $n=1,2,\dots$.
The degeneracy factor of each energy level is $2 n^2$. 
In addition to the bound states, the solution to the Schr\"odinger equation also includes scattering states corresponding to the energy of relative motion $E\ge 0$.

For $Z > 0$, in addition to the neutral atom in which $Z$ electrons are bound to the nucleus,
ions $Z_i$ can be formed, in which $Z-i$ electrons are bound to the nucleus, and the charge of the ion is $i\,e$, where $i=Z_i=0, 1,\dots, Z$.
First, we revisit the case of the zero-density limit, where no density effects occur (index “0”).
At zero density, the Schr\"odinger equation for the ion $Z_i$ describes the motion of the center of mass, which approximates the motion of the nucleus with charge $Ze$ and can be separated.
The intrinsic motion corresponds approximately to the motion of the $Z_i$ electrons in the Coulomb field of the nucleus at a fixed position,
\begin{eqnarray}\label{Sgl0}
   && \left(\frac{\hbar^2k_1^2}{2 m_e}+ \dots +\frac{\hbar^2 k_{Z-i}^2}{2 m_e}-E^{0,i+}_n\right)\psi^{0,i+}_n(k_1,\dots ,k_{Z-i})\nonumber \\
   && -\sum_{{\bf q}}\frac{Z\,e^2}{\epsilon_0 q^2}\sum_{l}^{Z-i} \psi^{0,i+}_n(k_1,\dots,k_{l}+{\bf q},\dots ,k_{Z-i}) \nonumber \\
   && +\sum_{{\bf q}}\frac{e^2}{\epsilon_0 q^2} \sum_{l<m}^{Z-i}\psi^{0,i+}_n(k_1,\dots,k_{l}+{\bf q},\dots ,k_m-{\bf q},\dots ,k_{Z-i})=0.
\end{eqnarray}
When there is more than one electron bound to a nucleus with proton number $Z$, the solution to the Schr\"odinger equation cannot be expressed analytically.
Approximation methods are used to determine the solution, such as the shell model, in which the interaction between the electrons is replaced by a mean field. 
For the wave function $\psi^{0,i+}_n(k_1,\dots ,k_{Z-i})$ of the ion with charge $+i\,e$, the antisymmetrized product ansatz (Slater determinant) of single-particle wave functions can be applied, whereby the mean field is described in a self-consistent manner.
More generally, density functional theory (DFT) was developed to obtain optimal single-particle wave functions.
Numerical methods such as path integral Monte Carlo (PIMC) can also be applied to find solutions that take correlations between the electrons into account.
We do not present here the success of atomic physics in solving Eq. (\ref{Sgl0}).
Instead, we use the empirical values of the energy levels $E^{0,i+}_n$ of the free ions, charge $Z_ie=i\, e$, excitation state $n$, and degeneracy factors $g^{i+}_n$, which can be found, for example, in the NIST tables \cite{NIST}.

\subsection{In-medium single-particle states}

We are interested in the influence of the surrounding plasma on the energy eigenvalues $E^{i +}_n(T, n_{\rm nuc})$, which depend on the matter density $n_{\rm nuc}$ and the temperature $T$ of the plasma.
The quantum statistical solution to this problem is found using the Green’s function method,
which introduces concepts such as the self-energy and the polarization function  \cite{KKER,R13}.
The perturbation expansion with respect to the Coulomb interaction can be represented by Feynman diagrams.

The single-particle propagator is determined by the self-energy, which depends on the wave number $k$ and the energy.
The energy of the free particle $c$ in the quantum state $k$, $E^{c,0}_k=\hbar^2 k^2/(2 m_c)$, is shifted by the self-energy, resulting in the quasiparticle energy.
Since the self-energy is generally energy-dependent, the shift $\Delta^{c,\rm qu}_k$ is the value of the self-energy at the quasiparticle energy $E^{c,{\rm qu}}_k=E_k^{c,0}+\Delta^{c,\rm qu}_k$.
In perturbation theory, the self-energy is expressed as the sum of terms represented by Feynman diagrams.

We consider an approximation of the self-energy given by the lowest-order Feynman diagrams. 
For the quasi-particle energy, we obtain the approximation
\begin{equation}
\label{SingleEk}
    E^{c,\rm qu}_k=E_k^{c,0}+\Delta^{c,\rm HF}_k+\Delta^{c,\rm scr}_k.
\end{equation}
The lowest-order contribution to the self-energy is the Hartree-Fock approximation. The Hartree term is zero due to charge neutrality. The Fock term is an exchange term that is relevant only for electrons that can be degenerate. It leads to a quasiparticle shift of the single-particle energies at wave number $\bf k$
\begin{equation}
\label{eq:HF}
\Delta_k^{e,\rm HF}=-\sum_{k'} V_{e,e}(k,k';k',k)\, f_e(k')
\end{equation}
with the Fermi distribution function of electrons
\begin{equation}
    f_e(k)=\frac{1}{\exp(E^{e,0}_k/k_BT-\mu^0_e/k_BT)+1}.
\end{equation} 
The electron chemical potential is given for the ideal (noninteracting) Fermi gas by
\begin{equation}
\label{nFermi0}
 n_e=\frac{1}{\pi^2}\int_0^\infty dk \frac{k^2}{\exp(E^{e,0}_k/k_BT-\mu^0_e/k_BT)+1}.
\end{equation}
The Fock term is a consequence of the Pauli principle, around an electron the electrons with the same spin orientation form the Fermi hole.
This effect can be neglected in classical plasmas but becomes important in degenerate plasmas, such as the electron gas in metals.

In the next order of perturbation theory, Debye screening emerges, which is already the dominant process in classical plasmas.
The screening in Coulomb plasmas is described by the polarization function $\Pi(q,\omega)$, which depends on the wave vector $\bf q$ and the frequency $\omega$.
In the simplest approximation, we obtain the Random Phase Approximation (RPA)
\begin{equation}
\label{RPA}
\Pi^{\rm RPA}(q,\omega)=\lim_{\epsilon \to 0}\sum_{c,k}\frac{f_c(k)-f_c({\bf k} + {\bf q})}{E^{c,0}_k-E^{c,0}_{{\bf k} +{\bf q}}-\omega+i \epsilon}
\end{equation}
which, in the static limit, leads to Debye screening of the Coulomb potential.
The statically screened interaction $V_{Q_1,Q_2}^{\rm scr}(q)$ is given by the Debye interaction in the low-density range, 
\begin{equation}
\label{eq:Debye}
   V^{\rm Deb}_{Q_1,Q_2}({\bf q})=\frac{Q_1Q_2}{\Omega_0\epsilon_0 (q^2+ \kappa^2)} 
\end{equation} 
with the Debye screening parameter ($n_i$ is the density of ions, charge $+i \, e$) 
\begin{equation}
   \kappa^2=\sum_i Z_i^2e^2 n_i/(\epsilon_0 k_BT)+\kappa^2_e, 
\end{equation}
\begin{equation}
\label{kappae}
 \kappa^2_e=\frac{8 \pi}{k_BT}  \left(\frac{2 \pi \hbar^2}{m_e k_BT}\right)^{-3/2} 
\frac{e^2}{4 \pi \epsilon_0}
\frac{1}{ \sqrt{\pi}} \int_0^\infty dt \frac{t^{-1/2}}{e^{t-\mu^0_e/k_BT}+1}.
\end{equation}
A point-like charge is assumed. Improvements to the Debye potential (12) that take into account higher orders of $q$ or the finite size of the nuclei are discussed in the literature; see, e.g., Ref. \cite{Barba}.

A charge $Z_ie=+i\,e$ immersed in the plasma generates a screening cloud. 
At low densities, the linearized Poisson-Boltzmann equation yields the Debye interaction.  
The expression for the Debye screening parameter $\kappa$ also includes the contribution of free electrons ($\kappa_e$), which may be degenerate. 
The electron contribution is then given by a Fermi integral, which, in the strongly degenerate limit, yields the Thomas-Fermi screening length instead of the Debye screening length; see the references \cite{OPAL,KKER,KSK}. 
The energy shift of a  particle with charge $Z_i e$ (including electrons, $Z_e=-1$) is 
\begin{equation}
\label{DebScr}
 \Delta^{i,\rm scr}_k \approx  \Delta_k^{i,\rm Debye}=-\frac{\kappa}{2}\frac{(Z_i e)^2}{4 \pi \epsilon_0} . 
\end{equation}

The Debye shifts are merely the lowest-order terms in the perturbation expansion of the polarization function, which is represented by Feynman diagrams.
The self-energy shifts have been improved; see \cite{Lin17}, where the ion structure factor is included.
Alternatively, semi-empirical interpolation expressions \cite{EK63,SP66} are introduced to extend the Debye approach to higher densities.
In the following, we consider the Stewart-Pyatt expression \cite{SP66}, which is frequently used in practical calculations; see the subsequent section \ref{sec:bound,scr} and Appendix \ref{sec:Carbon}.

\subsection{Bound states, density effects by screening of the interaction}
\label{sec:bound,scr}

Two-particle correlations are described by the two-particle Green's function, for which a Bethe–Salpeter equation can be derived.
If the interaction with the medium is neglected, the two-particle Green's function in the low-density limit is obtained by summing the ladder diagrams \cite{R13}
\begin{equation}
\label{G0}
G^0_2(k_1,k_2;k'_1,k_2';z)=\sum_{nP} \langle k_1,k_2|nP \rangle \frac{1}{E^0_{nP}-z} \langle nP: k_1',k_2' \rangle
\end{equation}
where $E^0_{nP}$ and $\psi^0_{nP}(k_1,k_2) = \langle k_1,k_2 | nP \rangle$ are the eigenvalues and eigenstates of the Schr\"odinger equation.
$P$ denotes the total momentum $\bf P$ and the total angular momentum, and $n$ is the intrinsic quantum number describing bound and scattering states. The analytical continuation from the Matsubara frequencies into the complex $z$-plane is considered \cite{KKER,R13}. A two-particle Schr\"odinger equation describing the interaction of an electron with an ion is given by Eq. (\ref{Sgl0}) for $i=Z-1$.

Similar Schr\"odinger equations can be formulated for multi-electron ions by considering the corresponding few-particle Green's functions.
In the low-density limit where in-medium effects can be neglected, the solution of the multi-electron Schr\"odinger equation (\ref{Sgl0}) 
provides the expression for the Green’s function that generalizes equation (\ref{G0}).

In the case of Carbon, for example, we have binding states for ions consisting of the nucleus and $6-i$ electrons, where the charge of the ions
takes the values $i = 0, 1, \dots, 5$.
Instead of solving the corresponding multi-electron Schr\"odinger equation, we can use the empirical values of the observed energy levels, which can be found, for example, in the NIST tables \cite{NIST}. 
The ionization energies $I^{0,i}_0$ are the minimum energies required to remove an electron from the ground state of the Carbon ion 
$C^{i+}$ (zero density): $I^{0,0+}_0 = 11.2603$ eV, $I^{0,1+}_0 = 24.38332$ eV, $I^{0,2+}_0 = 47.8878$ eV,
$I^{0,3+}_0 = 64.5$ eV, $I^{0,4+}_0 = 392.087$ eV, $I^{0,5+}_0 = 489.805$ eV. In addition, these tables also list the excitation energies of the various ions. 

We consider the modification of the multi-electron Schr\"odinger equation (\ref{Sgl0}) in a dense medium.
This quantum statistical problem is solved using the Green’s function method \cite{KKER,R13}.
We provide a simplified expression; a more detailed discussion can be found in \cite{RKKKZ78,ZKKKR78,KKER,KSK}.
We replace the dynamic screening with a statically screened interaction.
For an ion $i$ with charge $Z_i e$ and an electron, which form a cluster with charge $(i-1)e$, we obtain
\begin{eqnarray}
\label{BSE1}
&&\left[E^{e,{\rm qu}}_{k_1}+E^{i,{\rm qu}}_{k_2}
\right] \psi^{i-1}_{n}({\bf k_1},{\bf k_2})\nonumber\\&&
+[1-f_e(k_1)] \sum_{\bf q}V_{e,i}^{\rm scr}({\bf q}) \psi^{i-1}_{n}({\bf k_1}+{\bf q},{\bf k_2}-{\bf q})\nonumber \\
&&=E^{i-1,{\rm qu}}_{n,k_2} \psi^{i-1}_{n}({\bf k_1},{\bf k_2})
\end{eqnarray}
Here, $E^{c,{\rm qu}}_k = E^{c,{\rm qu}}_0 + \hbar^2 k^2/(2 m_c)$ if the shifts are not depending on $k$. Due to the large mass ratio $m_i/m_e \gg 1$, we set $k_2$ equal to the center of mass momentum, which can be separated.
The intrinsic motion follows from the Schr\"odinger equation in the medium
\begin{eqnarray}
\label{BSE2}
&&\left[E^{e,{\rm qu}}_{k}+E^{i,{\rm qu}}_{0}-E^{i-1,{\rm qu}}_{n}
\right] \psi^{i-1}_{n}({\bf k})\nonumber\\&&
+[1-f_e(k)] \sum_{\bf q}V_{e,i}^{\rm scr}({\bf q}) \psi^{i-1}_{n}({\bf k}+{\bf q})=0.
\end{eqnarray}
The statically screened interaction $V_{e,i}^{\rm scr}(q)$ is described in the low-density regime by the Debye interaction $V^{\rm Deb}_{e,i}(q)$ according to Eq. (\ref{eq:Debye}).

The polarization of the plasma by the charge $Q$ leads to a shift in the quasi-particle energies.
This part of the self-energy, which is due to correlations, is given at low densities by the Debye term
$-\kappa Q^2/(8 \pi \epsilon_0)$, which does not depend on $k$.
In the classical case, where $f_e(k) \ll 1$ so that degeneracy effects can be neglected, we expect the contribution to the ionization potential $I^{i-1}_0=E^{e,{\rm qu}}_0+E^{i,{\rm qu}} _0-E^{i-1,{\rm qu}}_0=I^{0,i-1}_0+\Delta I^{i-1}_0$ from the Debye shifts as
\begin{equation}
\label{DelIDebye}
    \Delta I^{i-1,\rm Debye}_{0}=\frac{\kappa}{2} \frac{e^2}{4 \pi \epsilon_0} [(i-1)^2-i^2-1]= -\frac{\kappa e^2}{4 \pi \epsilon_0} i.
\end{equation}
This result agrees with the solution of the in-medium Schr\"odinger equation (\ref{BSE2}), since, after expanding $e^{-\kappa r} \approx 1-\kappa r$, the Debye potential yields the contribution $-\frac{\kappa e^2}{4 \pi \epsilon_0} i$ to the energy $E_n^{i-1,{\rm qu}}$, in addition to $\Delta^{e,{\rm Debye}}+\Delta^{i,{\rm Debye}}$ according to the self-energy corrections.
This reduction in the ionization potential by the term $\Delta I^{i-1,\rm Debye}_{0}$ is the ionization potential depression (IPD) in the Debye approximation.

The Debye shifts are merely the lowest-order terms in the perturbation expansion of the polarization function, which is represented by Feynman diagrams.
The self-energy shifts have been refined; see \cite{Lin17}.
Alternatively, semi-empirical interpolation expressions are introduced to extend the Debye approximation to higher densities.
In practical calculations, the Stewart-Pyatt expression is frequently used, which interpolates between the Debye shift at low densities and the ion sphere model at high densities, 
where the ions are strongly correlated. The ionization potential depression (IPD) results from the 
energy shifts of the free-state continuum and the bound-state shift, such that the ionization energy is altered by the medium,  
\begin{equation}
\label{SPIPD}
 \Delta I^{i-1,\rm SP}_{0}=\frac{3}{2} \frac{(Z_i+1) e^2}{4 \pi \epsilon_0 r_{\rm IS}} 
\left[(1+s^3)^{2/3}-s^2 \right]
\end{equation}
with the ion sphere radius $r_{\rm IS}= (3 Z_i/(4 \pi n_e))^{1/3}$ and $s=(\kappa r_{\rm IS})^{-1}$.
In the present work, the Stewart-Pyatt expression (\ref{SPIPD}) was used instead of the Debye shift (\ref{DelIDebye}).

In the parameter range considered here, $\Delta^{i,\rm SP}_0$ depends only weakly on $T$ and $i$.
We use this expression for our calculations, App. \ref{sec:Carbon}. 
Within the framework of a more systematic quantum statistical approach, the Stewart-Pyatt expression \cite{SP66} 
for the IPD can be improved using the dynamical structure factor \cite{Lin17}.
There, a slightly stronger IPD is observed.

\subsection{Degeneracy Effects}

Compared to classical plasmas, the Pauli principle, which applies to fermions, gives rise to new effects when the plasmas are degenerate.
We use the relation (\ref{Theta}), where $n_e$ denotes the free electron density.

For single-particle states, the first-order contribution is the Fock shift; see Eq. (\ref{eq:HF}).
The Fock shift of an electron with momentum $\hbar {\bf k}$ is given by
\begin{equation}
\label{HF0}
\Delta^{e, \rm Fock}_k=-\sum_q \frac{e^2}{\Omega_0\epsilon_0 q^2}f_e({\bf k}+{\bf q}).
\end{equation}
In the limiting case of strong degeneracy, $T \ll T_{\rm F}$, we approximate the Fermi distribution function 
as a step function, $f_e(k )= \Theta(k_{\rm F}-k)$. The Fermi wave number is given by 
$k_{\rm F}=(3 \pi^2 n_e)^{1/3}$. 
The calculation of the Fock shift yields
\begin{equation}
\label{HF00}
\Delta^{e, \rm Fock}_k=-\frac{e^2}{4 \pi \epsilon_0} \frac{1}{\pi k} {\rm Re}\left[k\, k_{\rm F}+(k_{\rm F}^2-k^2)\arctan\left(\frac{k}{k_{\rm F}}\right)\right].
\end{equation}
If $k=0$, the following holds for the shift of the continuum edge:
$ \Delta^{e,\rm Fock}_0=
-\frac{2}{\pi} \frac{e^2}{4 \pi \epsilon_0} k_F$.
Whilst  screening makes the largest contribution to the single-particle shifts in classical plasmas,
the Fock shift dominates in the limiting case of strong degeneracy.

In degenerate plasmas, electrons in bound states are also subjected to the Pauli principle.
Here we consider the simplest case: an electron bound to a nucleus of charge $Z$.
The two-particle Schr\"odinger equation without medium corrections yields the well-known hydrogen-like solution for the ground state 
\begin{equation}
E_0=-Z^2 \frac{e^4}{(4 \pi \epsilon_0)^2} \frac{m_e}{2 \hbar^2}=-Z^2\,\, {\rm Ry}
\end{equation} 
and 
\begin{equation}
 \phi_0( k)= 8 \sqrt{\pi a_Z^3} \frac{1}{(1+a_Z^2 k^2)^2}, \qquad 
\psi_0(r)=\frac{1}{\sqrt{\pi a_Z^3}}e^{-r/a_Z}
\end{equation}
with $a_Z=\frac{4 \pi \epsilon_0}{Ze^2} \frac{\hbar^2}{m_e}$  (for instance, $a_C=1/6$ in Rydberg units).

At low densities, where the effects of the medium are weak, we can apply perturbation theory to calculate the in-medium shifts in the energies of all quasiparticles,
i.e., of the free components and of the bound states.
The shift in the bound state energy due to the Fock self-energy is then calculated using the unperturbed solution $\phi_0(k)$,
\begin{equation}
 \Delta E_0^{\rm Fock}=-\sum_{k,q}\phi^2_0( k) V_{ee}(q) f_e({\bf k}+{\bf q})
=-\frac{32}{\pi } \int_0^\infty dk \frac{k^2 a_Z^3}{(1+a_Z^2 k^2)^4}\Delta^{\rm Fock}_e(k )\,.
\end{equation}
The Pauli blocking shift at $T=0$ is given by (see Ref. \cite{IPD19})
\begin{equation}
\label{PauliT=0}
 \Delta E_0^{\rm Pauli}=-\sum_{k,q}\phi_0(k)f_e(k) V_{ei}(q) \phi_0({\bf k}+{\bf q})
=\frac{Z e^2}{4 \pi \epsilon_0} \frac{2}{\pi a_Z} 
\left[\frac{a_Z k_F (a_Z^2k_F^2-1)}{(a_Z^2k_F^2+1)^2}+\arctan(a_Z k_{\rm F})\right].
\end{equation}
For any $T$, we find
\begin{eqnarray}
    \Delta E_0^{\rm Pauli}&=&\frac{Z}{2 \pi^3}\int_0^\infty dk\,k\int_0^\infty dk'\,k'\phi_0(k) \phi_0(k') \nonumber \\ 
    && \times \frac{1}{2}[f_e(k)+f_e(k')]\ln \left[\frac{(k+ k')^2+\kappa^2}{(k- k')^2+\kappa^2}\right].
\end{eqnarray}
In the special case $\kappa \to 0$, we have    
\begin{eqnarray}
\label{Paulkap0}
     \Delta E_0^{\rm Pauli}&=& \frac{1}{2 \pi^2} \int_0^\infty dk \, k^2 \phi_0^2(k) (k^2+Z^2) f_e(k).
\end{eqnarray}
Similar expressions can be derived for the shifts of the excited states.
Further details can be found in Sect. \ref{sec:4}.

Bound states are also regarded as quasi-particles. This is referred to as the chemical picture.
In a quantum statistical approach, a corresponding propagator is introduced by equation (\ref{G0}). 
As a result of the interaction of the bound states with the plasma,
the energy eigenvalues are modified, $E_n = E^0_n + \Delta E_n$, where $\Delta E_n = \Delta E^{\rm Fock}_n + \Delta E^{\rm Pauli}_n$. The wave functions $\phi_n(\bf k)$ are also modified.
As the density increases, the influence of the plasma grows, and
the bound state energies can merge with the continuum, causing
bound states to disappear.
This dissolution of the bound states is known as the Mott effect and has important consequences for the macroscopic properties of the plasma;
further details can be found in \cite{RKKKZ78,ZKKKR78}.

\section{Composition}
\label{sec:3}

\subsection{The composition of the partially ionized plasma}

The total nuclear density $n_{\rm nucl}^{\rm tot}$ can be decomposed as follows (the ions are  non-degenerate):
\begin{equation}
\label{eq:cluster}
    n_{\rm nucl}^{\rm tot}=\frac{1}{\Omega_0}\sum_{i,n,p} e^{-\beta E^i_{n}(p) +\beta \mu_i}=\sum_{i=0}^Z n_i.
\end{equation}
In this cluster decomposition of the total density, the various channels are denoted by $i$ which is related to the number of particles; the center of mass momentum is $\hbar {\bf p}$; and $n$ denotes the intrinsic state of the cluster, such as excited bound states. 
For example, $i=0, 1, \dots,Z$ represents the ionization state and describes the number of particles (the nucleus and $Z-i$ electrons) for which correlations are taken into account. 
The sum over the spin orientation and the total angular momentum yields a degeneracy factor. 
The intrinsic quantum state $n$ encompasses not only the excited bound states, but also the continuum states.

For each component of the plasma, a corresponding chemical potential $\mu_i$ can be introduced.
In chemical equilibrium, $\mu_i$ does not depend on $n$, and in equation (\ref{eq:cluster}) we have
\begin{equation}
    \mu_i=\mu_{\rm nucl}+(Z-i) \mu_e.
\end{equation} 
The chemical potentials $\mu_{\rm nucl}, \mu_e$ of the elementary constituents are in turn related to each other by the condition of neutrality. It should be noted that larger clusters $i > Z$ can also be included, such as the formation of H$^-$ or molecules; however, this would go beyond the scope of this work.

Since the ions can be regarded as classical particles, the sum/integral over $\bf p$ is simply evaluated, yielding the following result (where $\Omega_0$ is the volume of the system):
\begin{equation}
    \frac{1}{\Omega_0}\sum_p e^{\hbar^2 p^2/(2 m_Z k_BT)}=\int\frac{d^3p}{(2 \pi)^3}e^{\hbar^2 p^2/(2 m_Z k_BT)}=\left( \frac{m_Zk_BT}{2 \pi \hbar^2} \right)^{3/2}=\Lambda^{-3}_{\rm nucl},
\end{equation}
the mass $m_Z$ of the various ions is nearly identical.
The relation (\ref{eq:cluster}) can be rewritten as
\begin{equation}
\label{eq:cluster1}
    n_{\rm nucl}^{\rm tot}=\Lambda^{-3}_{\rm nucl}\sum_{i=0}^Z e^{\beta \mu_i} \sigma^i, \qquad  n_{i}=\Lambda^{-3}_{\rm nucl}e^{\beta \mu_i} \sigma^i
\end{equation}
with the intrinsic partition function
\begin{equation}
\label{eq:sig0}
    \sigma^{0,i}=\sum_n e^{-\beta E^i_{n}}=e^{-\beta E^i_{0}}\sum_n e^{-\beta E^{i,*}_{n}}
\end{equation}
where $E^{i,*}_{n}=E^i_{n}-E^i_0$ denotes the excitation energies. For the free ions, these excitation energies are found in tables, e.g. \cite{NIST}, together with the degeneracy factors according to the quantum numbers of angular momentum  (including spin).

A further rewriting is used in the literature, introducing the ionization potentials $I_0^i$.
The ionization potential is the minimum energy required to extract an electron from the ion $i$ in the ground state, energy $E^i_{0}(p=0)$. The final state after ionization has the energy  $E^{i+1}_{0}(p=0)+E^e_k$
where $\hbar {\bf k} $ gives the momentum of the extracted electron. If $k=0$ has the lowest possible energy of the emitted electron, the edge of the continuum gives the ionization potential
\begin{equation}
    I^i_0=E^{i+1}_{0}(0)+E^e_0-E^i_{0}(0).
\end{equation}
Values for the free components are found in the tables, e.g. \cite{NIST}. 
In a dense medium, we have to use the quasiparticle energies. 
For degenerate electrons at zero temperature, only final states above the Fermi energy are possible.

We can replace in Eq. (\ref{eq:cluster1}) 
\begin{equation}
    E^i_{0}=E_{\rm nucl}+(Z-i) E^e_0-\sum_{i'=0}^{Z-i} I^{Z-i'+1}_0,
\end{equation}
where the lowest energies of the free components, $E_{\rm nucl}$ and $E^e_0$, are taken as zero in our calculations.
To eliminate both $\mu_{\rm nucl}$ and $\mu_e$, a second equation for $n_e^{\rm tot}$ is required.
Similar to Eq. (\ref{eq:cluster}), we have
\begin{equation}
\label{eq:clustere}
    n_{e}^{\rm tot}=\sum_{i=0}^Z (Z-i) n_i +n_{e}.
\end{equation}
Part of electrons are localized within the ions, the other part moves freely.
Since the electron mass is small relatively to the ion mass, the free electrons are described by quantum statistics which gives at arbitrary degeneracy the Fermi distribution,
\begin{equation}
\label{nFermi}
    n_{e}=\frac{1}{\Omega_0}\sum_{k}f^{\rm qu}_e(k)=2\int \frac{d^3k}{(2 \pi)^3}\frac{1}{e^{\beta (E^{e,\rm qu}_k-\mu_e)}+1}
\end{equation}
(spin factor 2). Coming back to Eq. (\ref{nFermi0}), which relates the density to the chemical potential in the ideal Fermi gas approximation,
the relation (\ref{nFermi}) takes interaction into account. 
This expression for the free electron density determines also screening, $n_{e}^{\rm free}/n_e^{\rm total}$ is the ionization degree.
The densities of the different components of the plasma are related by the neutrality condition
$\sum_i Z_i n_i=\sum_i i\, n_i=n_{e}$. 

\subsection{The intrinsic partition function $\sigma^i(n,T)$}

The intrinsic partition function (\ref{eq:sig0}) is obtained if the interaction between the components of the plasma, the free and bound states, is neglected. 
However, an ideal, noninteracting plasma with occasional reactions to establish chemical equilibrium
is not appropriate for a dense plasma where interactions have to be taken into account.
This is consistent with the account of excited states which include also scattering states.
We are confronted with two problems: (i) the treatment of in-medium effects which leads to the quasiparticle concept, the in-medium shift of energy levels and medium-modified scattering phase shifts, and (ii) the subdivision into free and correlated density of the electrons.

Solving these problems requires a quantum statistical approach where well-defined quantities such as the spectral function are introduced.
We give here only the main ideas.
To discuss the long-standing problem of the subdivision of the electron density into a bound part and a free part,
we consider the  second virial coefficient for a short-range potential.
An exact expression is given by the Beth-Uhlenbeck formula, for a given channel $\alpha$ (spin, angular momentum, etc.)
we have
\begin{eqnarray}
\label{B-U}
&&\sigma^{i}_{\alpha}(T)=\sum_n^{\rm bound}   e^{-E_{\alpha,n}/k_BT}+\int_0^\infty \frac{dE}{\pi} e^{-E/k_BT}\frac{d}{d E} \delta_\alpha(E) 
\end{eqnarray} 
where $\delta_\alpha(E)$ is the scattering phase shift in the channel $\alpha$. 
Both, $E_{\alpha,n}$ and $\delta_\alpha(E)$, can be determined from experiments.
After integration by parts and using the Levinson theorem, as well as introducing the quasiparticle energies, we obtain
\begin{eqnarray}
\label{n2virial}
&&\sigma^{i}_{\alpha}(T)= 
 \sum_n^{\rm bound}   (e^{-E^{\rm qu}_{\alpha, n}/k_BT}-1)+\int_0^\infty \frac{dE}{\pi k_BT} e^{-E/k_BT}\left\{ \delta_\alpha(E)-\frac{1}{2} \sin [2  \delta_\alpha(E)]\right\}. 
\end{eqnarray}
In this second version,\\ (i) we can assume that the contributions of the continuum remain small and can be neglected. 
We use this relation to define $\sigma^{i}(T)= \sum_n  (e^{-E_n/k_BT}-1)$ as the contribution of correlations (bound states).\\
(ii) An additional sin-term appears in the contribution of scattering states. 
This term is related to the change in the single-particle contribution (\ref{nFermi}) where quasiparticles are considered. 
The quasiparticle shifts such as the Hartree-Fock shift contain already part of the interaction, this must be extracted from the second virial coefficient to avoid double counting.\\
(iii) Also for the two-particle properties, bound state energies and scattering phase shifts, the in-medium corrections have to be considered, in particular the quasiparticle energy shifts of the bound states.

In the case of Coulomb interaction, scattering phase shifts can not be defined in the standard way because of the long-range character of the interaction. 
This problem has been investigated in plasma physics since a long time, see Refs. \cite{OPAL,KKER,KSK}. 
Already in the approximation where the interaction between the components of the partially ionized plasma is neglected,
the definition of the intrinsic partition function $\sigma_i^0(T)$, Eq. (\ref{eq:sig0}), contains problems as described by Planck, Larkin, and others \cite{Planck,Larkin,Larkin1}.
The sum over all bound states for the Coulomb system is divergent, and truncation must be performed to obtain convergent results.

For the Hydrogen atom, a convergent expression was proposed by Planck and Larkin  \cite{Planck,Larkin,Larkin1}
\begin{equation}
\label{PLC5}
\sigma^i(T)= e^{-\beta (E^{i+1}_0+E^e_0)}
\sum^{\rm bound}_n \left[e^{\beta I^{i}_{n}}-1-\beta I^{i}_{n}\right]
\end{equation}
where $n$ covers the intrinsic quantum numbers (including spin) of all bound states, $I^{i}_n=E^{i+1}_0+E^e_0-E^i_n$.
For free electrons, the state with lowest energy is $k=0$ so that $E^e_0=0$. 
For H, $i=0$, is H$^{1+}$ the free nucleus which has also the minimum energy $E^{1+}_0=0$ so that $I^{0}_n=1/n^2$.
For He, $i=0$, is He$^{1+}$ the He ion with one electron, and the minimum energy is $E^{1+}_0=-4$ Ry.
The additional last term in the partition function (\ref{PLC5}) is related to the Debye shift of the quasiparticle energy which is necessary for Coulomb systems,
it changes the analytical form of the virial expansion \cite{RLER26,Rpop26}, and terms of the form $n_e^{1/2}$ appear.
The case of Hydrogen is widely discussed in the literature, see Ref. \cite{Ebeling26} and further references given there. 

The remaining part of the continuum of scattering states may be neglected. 
Note that the subdivision of the intrinsic partition function into a bound state part 
and a scattering part is not free of ambiguity so that the definition of the ionization degree 
is model dependent. 
One possibility for linking the concept of free electron density to physical properties is based on the screening of the Coulomb interaction potential in plasmas \cite{Rpop26,RLER26}.
The polarisation function $\Pi(q,\omega)$ can provide us with a free electron density defined by its behavior at $\omega=0$ and
$q \to 0$.

\section{The single bound electron problem}
\label{sec:4}

We investigate the motion of an electron in the screened potential of an ion with charge $Ze$.
This model describes the final stage of ionisation in a plasma at increasing density; it defines the transition to a fully ionised plasma.
Furthermore, it also describes the mean-field approximation, in which the potential of the nucleus is replaced by an effective potential of the ion.
This model is suitable, for example, for a valence electron on top of filled shells.
The problem of two electrons moving in the mean-field potential is discussed in Sect. \ref{sec:5}.

As explained in Sect. \ref{sec:2}, within the framework of a quantum statistical approach, the  in-medium  Schr\"odinger equation is given by (\ref{BSE1}).
As usual, after separation of the momentum of the center of mass motion (which practically coincides with the momentum of the ion) the relative motion remains, which describes the intrinsic structure, i.e., the bound and scattering states, of the electron-ion system.
Energy levels and wave functions result from the solution of the corresponding  in-medium Schrödinger equation for the relative motion (\ref{BSE2}),
\begin{equation}
\label{BSE3}
    \left[E_k^{e,\rm qu}+E_{0}^{i+1,\rm qu} -E^{i,\rm qu}_{n}\right]\psi_{n}^i({\bf k})-\left[1-f_e(k)\right]\int \frac{d^3 q}{(2 \pi)^3}\frac{8 \pi Z} {q^2
    +\kappa^2}\psi_{n}^i({\bf k}+{\bf q})=0.
\end{equation}
The quasi-particle energy $E_k^{e,\rm qu}$ (\ref{SingleEk}) includes the Fock shift $\Delta^{e, \rm HF}_k$, Eq. (\ref{HF0}), and the shift $\Delta^{e,\rm scr}_k$ due to screening. Dynamic screening has already been considered in the literature; see \cite{RKKKZ78,ZKKKR78}; for simplicity, we assume static screening here, which leads to the Debye term in Eq. (\ref{DebScr}).
In this section, we use Ry units.

The edge of the continuum states $E^{i,\rm qu}_{{\rm cont}}$ from Eq. (\ref{BSE3}) is given by $k=0$: $E^i,{\rm qu}_{{\rm cont}}=E_{0}^i+1,{\rm qu}+\Delta^{e,\rm HF}_0+\Delta^{e,\rm scr}_0$. For a given $T$, the edge of the continuum states shifts downward as the density increases. Expressions for the Debye shift of the ion, charge $(Z_i+1)e$, and the electron are given in Eq. (\ref{DebScr}). The Fock term follows from Eq. (\ref{HF0}) for $k \to 0$.

In the non-degenerate case, we consider the screening terms.
Since the Debye shifts do not depend on $k$, they can be combined, and the difference 
\begin{equation}
E_0^{e,\rm qu}+E_{0}^{i+1,\rm qu} -E^{i,\rm qu}_{n} =I^i_{n}
\end{equation}
is the ionization potential of the state $\psi_{n}^i({\bf k})$. 
In this classical case, the ionization potential depression is given by the Debye shifts, which can be refined using the Stewart-Pyatt expression (\ref{SPIPD}).

In the degenerate case, the Fock self-energy term $\Delta^{e,\rm HF}_0$ appears in the shift of the continuum edge, but $\Delta^{e,\rm HF}_k$ also appears in the calculation of the bound states. If we assume that the $k$-dependence can be neglected, we can rewrite equation (\ref{BSE3}) in the form
\begin{equation}
\label{BSE4}
    \left[k^2+I^i_{n}\right]\psi_{n}^i({\bf k})-\left[1-f_e(k)\right]\int \frac{d^3 q}{(2 \pi)^3}\frac{8 \pi Z} {q^2
    +\kappa^2}\psi_{n}^i({\bf k}+{\bf q})=0.
\end{equation}
As explained above, the appearance of $\kappa$ in the interaction without the Pauli blocking factor $\left[1-f_e(k)\right]$ already describes the classical part of the IPD, which is enhanced by the Pauli blocking terms in the degenerate case.
Neglecting the $k$-dependence of the Fock shift is justified near the Mott point, where the wavefunction approximates that of a free electron.
Although it is possible to account for the $k$-dependence, it has only a minor effect on the IPD; see Ref. \cite{IPD19}.

The integral equation (\ref{BSE4}) is not Hermitean but can be symmetrized introducing $\psi_{n}^i({\bf k})/(1-f_e(k))^{1/2}=\phi_{n}^i({\bf k})$.
We can assume that the ground state is spherical symmetric ($s$ wave, c.m. momentum 0) so that
\begin{equation}
    \left[k^2+I^i_{n}\right]\phi_{n}^i(k)-(1-f_e(k))^{1/2}\int \frac{d^3k'}{(2 \pi)^3} \frac{8 \pi Z}{({\bf k}-{\bf k'})^2+\kappa^2} (1-f_e(p'))^{1/2}\phi_{n}^i(k')=0.
\end{equation}
Performing the angular integral, we obtain
\begin{equation}
    \left[k^2+I^i_{n}\right]\phi_{n}^i(k)-(1-f_e(k))^{1/2} \frac{Z}{\pi} \int_0^\infty dk' \frac{k'}{k}\ln \left[\frac{(k+ k')^2+\kappa^2}{(k- k')^2+\kappa^2}\right] (1-f_e(k'))^{1/2}\phi_{n}^i(k')=0.
\end{equation}
or, with $\phi_{n}^i(k)=u_{n}^i(k)/k$, the symmetric form
\begin{equation}
\label{Eq:ui}
\left[k^2+I^i_{n}\right]u_{n}^i(k)-(1-f_e(k))^{1/2} \frac{Z}{\pi} \int_0^\infty dk' \ln \left[\frac{(k+ k')^2+\kappa^2}{(k- k')^2+\kappa^2}\right] (1-f_e(k'))^{1/2}u_{n}^i(k')=0.
\end{equation}


We present results for $Z=6$, carbon, and calculate the energy shifts for the final ionization process, the reaction C$^{5+}\leftrightharpoons {\rm C}^{6+}+e$. The disappearance of the ionization potential determines the emergence of the fully ionized carbon plasma. 
We begin with the question:
How good is perturbation theory? In particular, is there an exact density value (the Mott density) at which the bound states disappear?

To illustrate this, let us consider the case of $C^{5+}$. The Debye shift of the continuum ($C^{6+} + e$) is $\Delta^{\rm cont} = -\kappa Z_{\rm C} e^2/(4 \pi \epsilon_0)$, see Eq. (\ref{DelIDebye}). 
This result can be refined using the Stewart-Pyatt expression or other results.
If we consider the Debye interaction $-Z_{\rm C} e^2/(4 \pi \epsilon_0 r)\exp(-\kappa r) \approx -Z_{\rm C} e^2/(4 \pi \epsilon_0 r) (1-\kappa r)$, perturbation theory yields a shift in the bound state energy of $+Z_{\rm C} e^2/(4 \pi \epsilon_0) \kappa$. 
This compensation effect was found in \cite{RKKKZ78,ZKKKR78}. 
In the case of H, the bound state energy of the neutral H atom remains unchanged, while the bound state energy of C$^{5+}$ and the continuum exhibit the additional Debye shift $-(Z_{\rm C}-1)\kappa e^2/(4 \pi \epsilon_0)$. 
Therefore, the IPD of C$^{5+}$ is mainly determined by the medium-induced modification of the interaction term, i.e., by dynamic screening and Pauli blocking.
The situation is illustrated in Fig. 1 of Ref. \cite{IPD19}.

We are concerned with the solution of Eq. (\ref{Eq:ui}) or, after discretization,
\begin{equation}
\label{Eq:ui1}
    \sum_\beta \left[k_\alpha^2 \delta_{\alpha\beta}-(1-f_e(k_\alpha))^{1/2} \frac{Z\,d}{\pi}  \ln \left[\frac{(k_\alpha+ k_\beta)^2+\kappa^2}{(k_\alpha- k_\beta)^2+\kappa^2}\right] (1-f_e(k_\beta))^{1/2}\right] u_n^i(k_\beta)=-I^i_{n}u_{n}^i(k_]\alpha),
\end{equation}
with $k_\alpha=d(\alpha-1/2), \alpha=1,2,\dots N_\alpha$, and $\alpha \longleftrightarrow \beta$. 
The limit $N_\alpha \to \infty, d \to 0$ must be calculated until convergence is achieved.

It is interesting to compare the exact solution of the in-medium Schr\"odinger equation with perturbation theory.
The perturbation approach based on the undisturbed wave function is not valid near the Mott transition; the wave function is strongly distorted due to medium effects. The Mott transition requires an exact description, regardless of whether the transition is sharp or an adiabatic approximation to the continuum states is present.

\begin{figure}[h]
  \centering
  \includegraphics[width=0.7\linewidth]{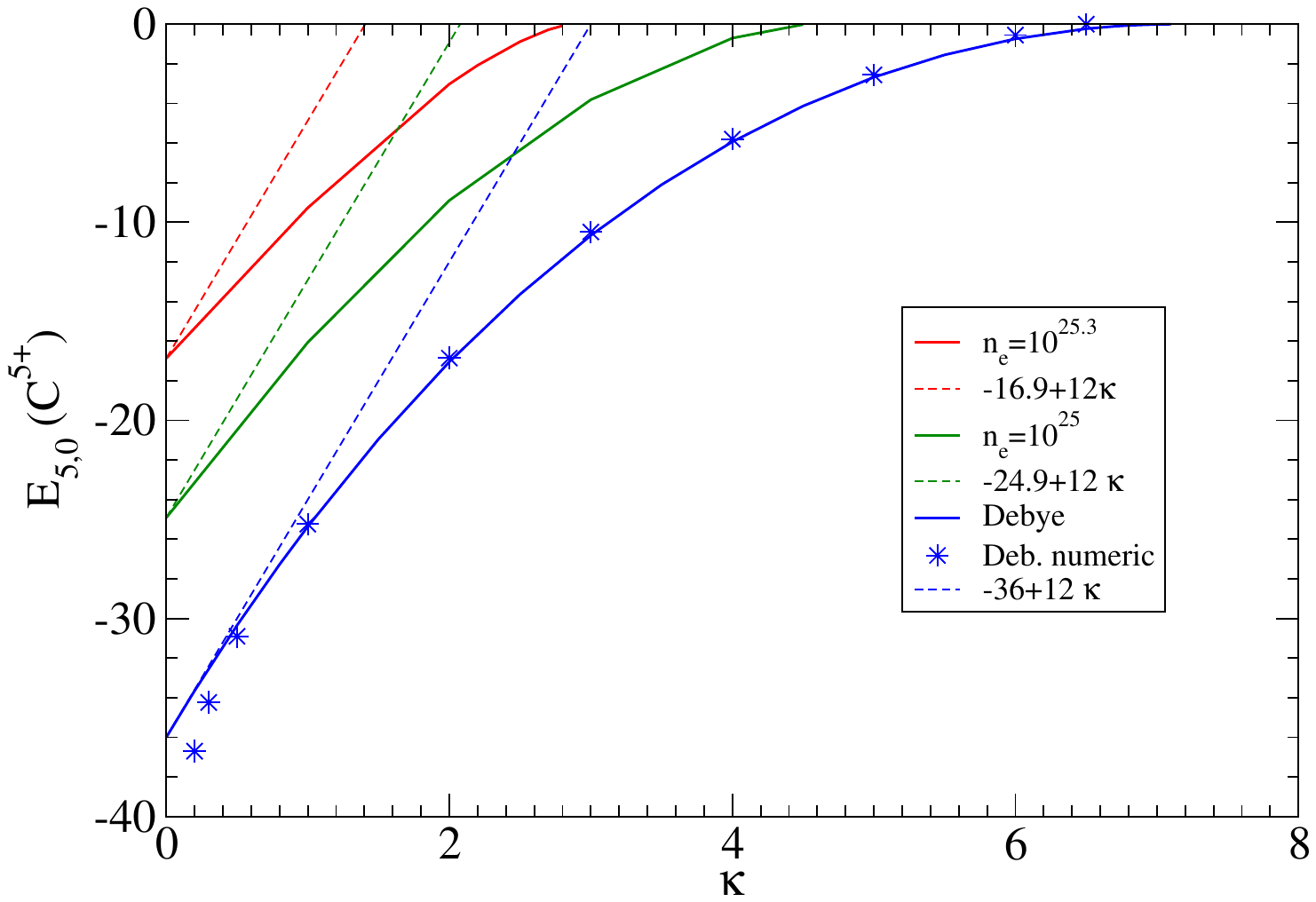}
   \caption{In-medium shift of the ground state energy $E_{\rm bound}=-I^{5+}_{0}$ of C$^{5+}$ as function of the screening parameter $\kappa$. 
   Numerical solution compared with perturbation theory (Ry units). \label{fig:DebyeC5}
   }
\end{figure}

The calculations are shown in Fig. \ref{fig:DebyeC5}.
For the Debye potential $-2 Z/r \exp(-\kappa r)$ (Ry units), the lowest-order perturbation theory yields a shift of $2 Z \kappa$. 
The exact solution yields a Mott point at $\kappa_{\rm Mott}=7.1$ instead of the value 3 from perturbation theory.
However, there is a sharp transition at which the bound state disappears and a resonance in the continuum appears.

Numerical calculations using Eq. (\ref{Eq:ui1}) were performed for $N_\alpha \le 30$ and various values of $d$, with the optimal value lying in the range $d=1.5$. 
The results can be improved by considering larger values of $N_\alpha$.
To assess the accuracy, we first present results for various $\kappa$ values for C$^{5+}$. 
In this work, we have not investigated the convergence of our calculations and therefore cannot provide error bars for the calculations.
We can only highlight the general trends.

We treat $\kappa$ as a parameter and solve the corresponding Schr\"odinger equation without the Pauli blocking terms.
Perturbation theory yields the shift $\kappa Z e^2/(4 \pi \epsilon_0)$ for the Debye potential, while the solution to the Schrödinger equation exhibits a deviation that is also approximated by the numerical solution. Deviations of the numerical solution from the solution of the differential equation occur for small $\kappa$, since the logarithm in Eq. (\ref{Eq:ui1}) becomes very sharp and must be resolved.

Furthermore, Fig. \ref{fig:DebyeC5} illustrates the effect of Pauli blocking. We solve the Schrödinger equation with the Pauli-blocking terms for a given free electron density and $T=100$ eV, which determines the Fermi functions, treating $\kappa$ as a variable.
The perturbation theory based on the free bound state  does not yield useful
shifts, since the ground state is already modified by Pauli blocking.

\begin{figure}[h]
  \centering
  \includegraphics[width=0.7\linewidth]{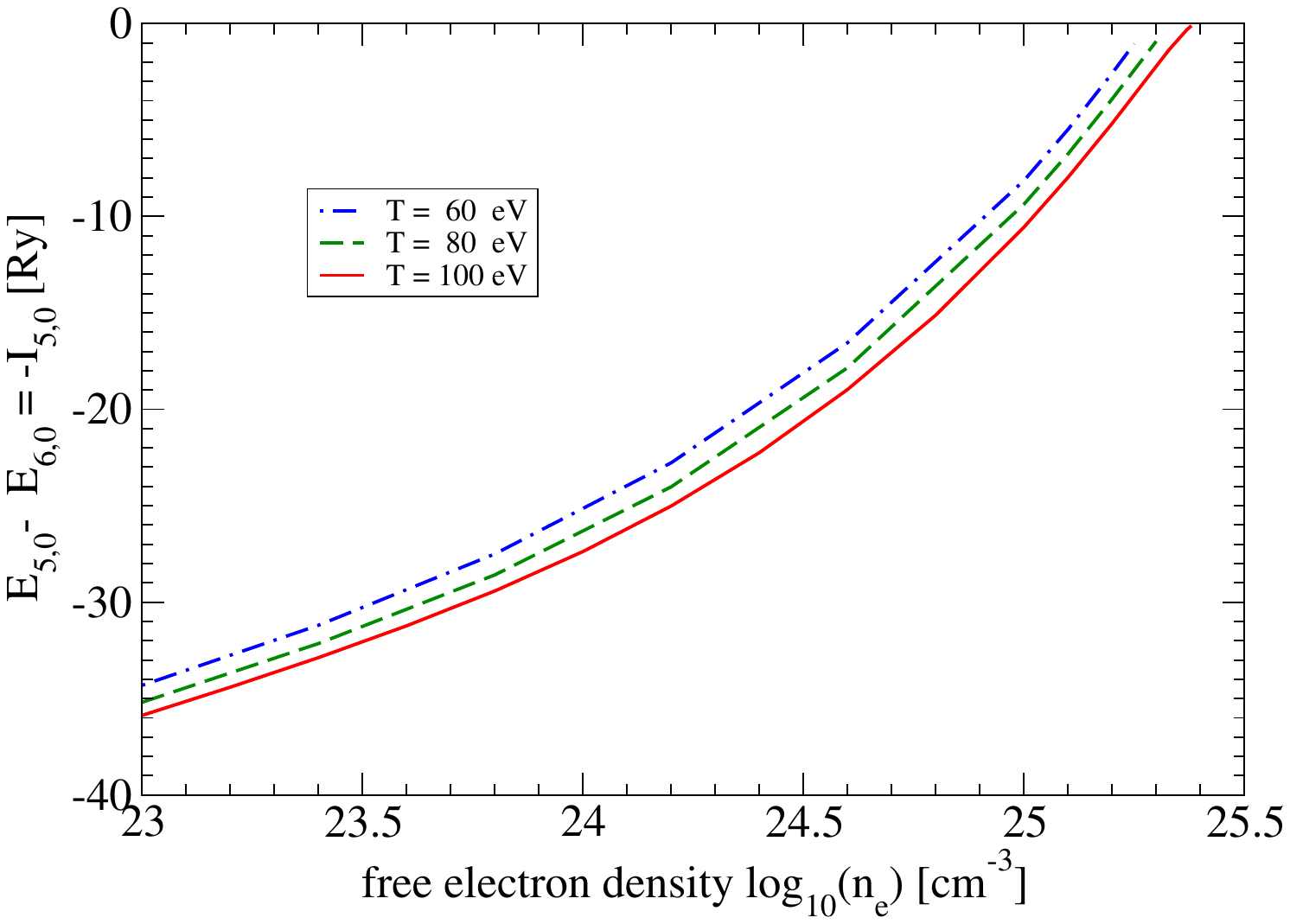}
   \caption{Ionization potential $-I^{5+}_{0}$ of C$^{5+}$ as function of density for different temperatures. \label{fig:IC5}
   }
\end{figure}

We present consistent calculations for the IPD $I_0^{5+}$ of C-plasmas, in which, for given values of $T$ and $n_e$ of free electrons, the Fermi functions according to Eq. (\ref{nFermi0}) ($\mu_e$) and the screening parameter $\kappa$ according to Eq. (\ref{kappae}) are calculated.
The results are shown in Fig. \ref{fig:IC5}, where various temperatures were considered.
The abrupt disappearance of the bound states is evident.
The dependence on $T$ is only weak.

The screening of the Coulomb interaction is a dynamic process; see Eq. (\ref{RPA}) for the polarisation function.
The use of a statically screened Debye potential in the in-medium Schr\"odinger equation (\ref{BSE3}) to describe the contribution of the ions is not justified at high densities. In particular, the ions cannot follow the rapid motion of the electrons. 
The ions are strongly correlated, so they can make only a small contribution to the screening.
Therefore, in our calculations we consider only the electron component of the screening.
To include the ion contributions, an improved treatment based on dynamic screening is required.

It should be noted that the solution to Eq. (\ref{Eq:ui1}) yields  also both excited states and continuum states.
The contribution of resonances can be determined from the density of states.
According to the generalized Beth-Uhlenbeck formula, we can separate the contribution of the single quasiparticles 
from the correlated part, which contains both bound states and correlations in the continuum.
The contribution of continuum correlations can become significant at low temperatures, where the Fermi function can be replaced by the step function.
This also solves the problem of the increase in ionization energy at $T=0$, since all states below the Fermi energy are occupied.

\begin{figure}[h]
  \centering
  \includegraphics[width=0.7\linewidth]{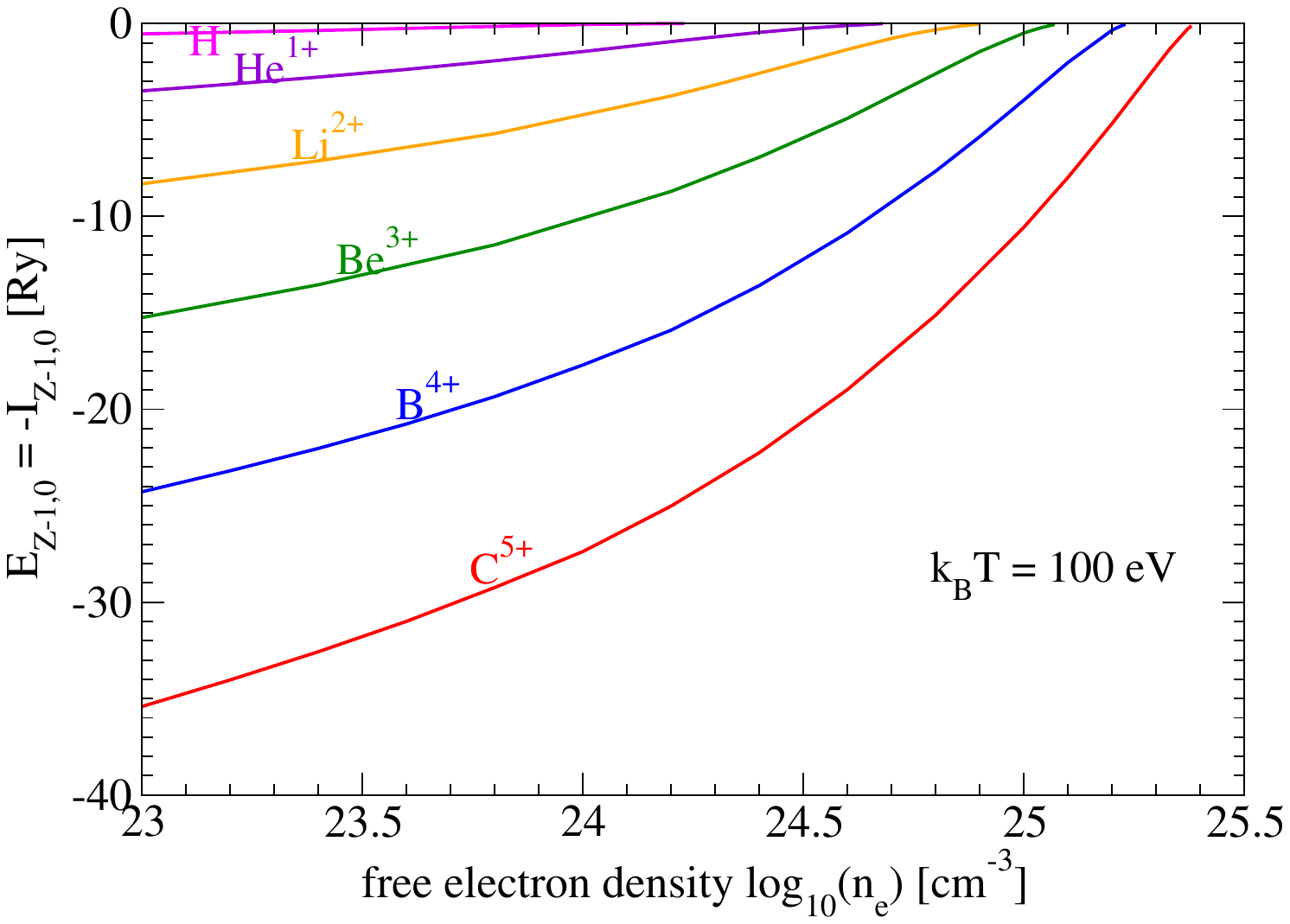}
   \caption{Ionization potential as function of density for different effective charge of the nucleus, temperature $k_BT=100$ eV. 
   \label{fig:ICnu}
   }
\end{figure}

Calculations were also performed for other materials with $Z \le 6$. The ionization potentials for the hydrogen-like bound states are shown in Fig. \ref{fig:ICnu} as a function of the free electron density at $k_BT=100$ eV. The bond states dissolve due to screening and Pauli blocking
at Mott densities; see Table \ref{Tab:Mott}.

\begin{table}[h]
\begin{center}
\caption{Mott densities at $k_BT=100$ eV for different materials, proton number $Z$ \label{Tab:Mott}}
 \begin{tabular}{|c|c|c|c|c|c|c|}
\hline
$Z$& 1 [H] & 2 [He]& 3 [Li]& 4 [Be]& 5 [B]& 6 [C]\\
\hline
$n^{\rm Mott}_e$ / [$10^{24}$ cm$^{-3}]$ & 1.74 & 4.79 & 7.94 & 11.7 & 16.9 & 23.4 \\
\hline
 \end{tabular}
\end{center}
\end{table}

\section{Two-electron ions}
\label{sec:5}

The multi-electron Schr\"odinger equation (\ref{Sgl0}) for an ion embedded in a plasma includes medium corrections, the self-energy shifts of the single-particle energies, and the Pauli blocking terms of the interaction. Standard solutions for this few-particle problem introduce an mean potential in which the Coulomb potential of the nucleus is screened by the average density of the other electrons.
The total wave function of the electrons is approximated by the antisymmetrised product of single-electron states.
We discuss the case of two electrons bound to a nucleus, i.e. He-like ions.
To estimate the effect of Pauli blocking, we consider the wave function in momentum space, where the occupation of phase space by the free electrons in the plasma becomes apparent.

We begin with the isolated neutral helium atom, i.e., with vanishing plasma density. 
Two electrons are bound to a nucleus with $Z=2$; the ground-state energy is  -5.80677 Ry \cite{NIST}.
Within the framework of a variational approach, we consider the  product ansatz (spin singlet) for the wave function
\begin{equation}
\label{phiHe}
    \psi_0 (r_1,r_2)=\frac{1}{\pi a^{3} }e^{-r_1/a}e^{-r_2/a}
\end{equation}
With the Hamilton operator, see Eq. (\ref{Sgl0}), the total energy is calculated as
\begin{equation}
    E_0=\frac{\hbar^2}{m_e}\frac{1}{a^2}-\frac{e^2}{4 \pi \epsilon_0} \frac{27}{8 a}.
\end{equation}
It reaches its minimum value of $-3^6/2^7\,{\rm Ry}=-5.6953$ Ry for $a=16/27\, a_B=0.5926\, a_B$.
This estimate of the binding state energy is slightly higher than the empirical bound state energy of -5.80677 Ry. 
The difference is due to the simple form of the wave function and the neglect of correlation effects; see Hylleraas \cite{Hylleraas}.

The effective potential contains the Coulomb potential of the nucleus, charge $Z$, which is screened by the other electron,
\begin{equation}
    V_1^{\rm eff}(r_1)=-\frac{Ze^2}{4 \pi \epsilon_0 r_1}+\frac{1}{2}\frac{e^2}{4 \pi \epsilon_0}\int d^3r_2 \psi_2^*(r_2)\frac{1}{|{\bf r}_1-{\bf r}_2|}\psi_2(r_2)\approx -\frac{Z_1^{\rm eff}e^2}{4 \pi \epsilon_0 r_1}.
\end{equation}
The use on an effective charge $Z_1^{\rm eff}=27/16=1.6875$ allows to use Hydrogen like orbitals.
It reproduces the wave function determined by the minimum value of the bound state energy, but it also reproduces the value of the bound state energy.

To investigate the influence of the plasma, we consider the case of strong degeneracy, in which the leading terms are the Fock self-energy shift and the Pauli blocking.
Since the self-energy shift of a weakly bound electron and a free electron is almost equal, the dissolution of the bound state is mainly due to Pauli blocking.
To estimate the shift of the ground state energy, we use the mean-field approximation, in which both electrons move independently on hydrogen-like orbitals.
Pauli blocking means that the part of the orbitals, which are already occupied by the free electron states of the plasma, cannot be accessed.
The corresponding shift in the bound state energy is given by Eq. (\ref{Paulkap0}).
In the strongly degenerate case, where the Fermi function can be replaced by the step function, we obtain the expression (\ref{PauliT=0}).

\begin{figure}[h]
  \centering
  \includegraphics[width=0.7\linewidth]{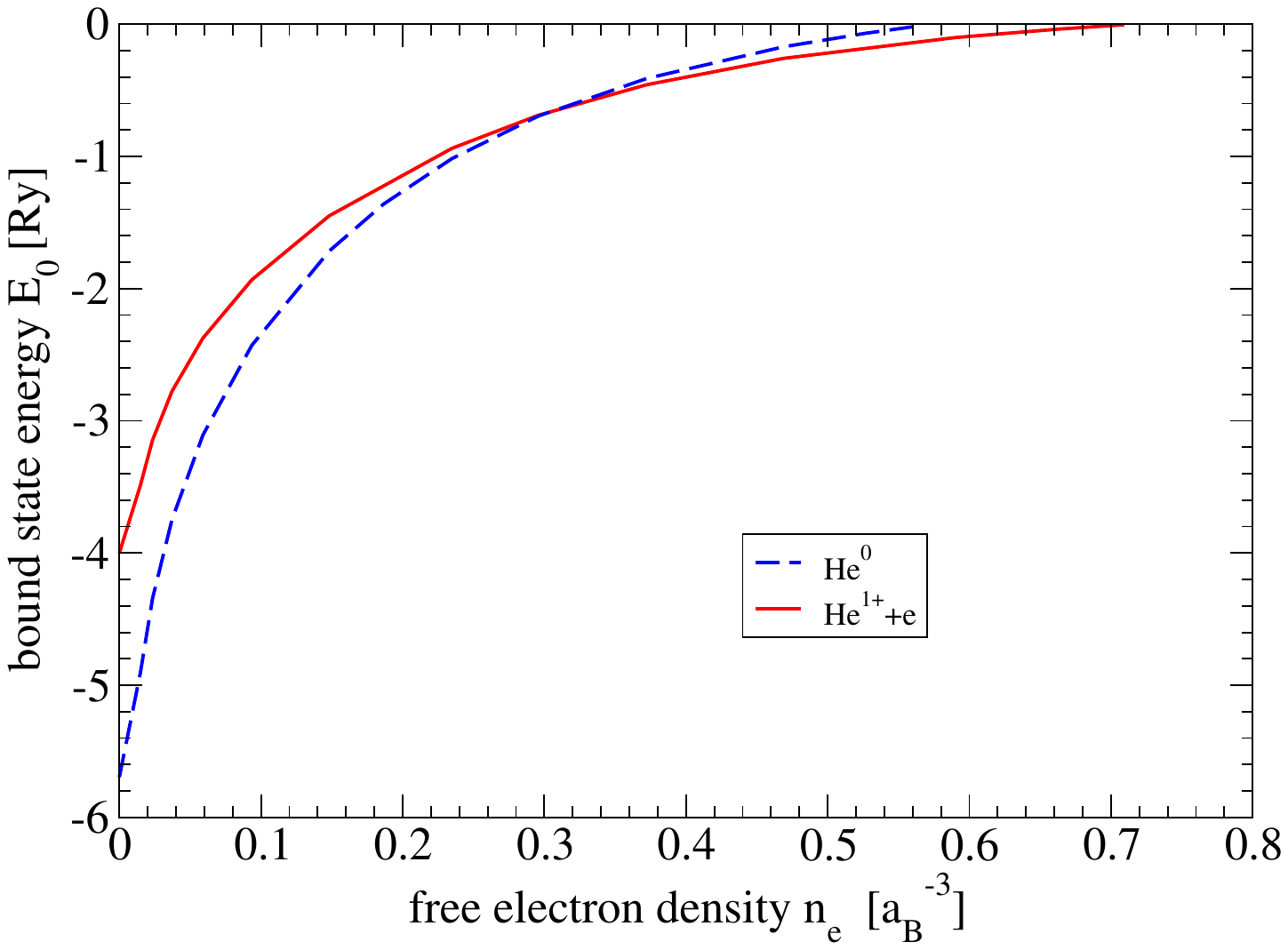}
   \caption{Bound state energy of He$^{0+}$ as function of free electron density, temperature $k_BT=100$ eV.
   For comparison, the bound state energy of He$^{1+}$ is also shown. 
    \label{fig:He+0}
   }
\end{figure}

A calculation for $k_BT=100$ eV is shown in Fig.  \ref{fig:He+0}. The bound state energy of He$^0$ at $n_e=0$ is -5.6953 Ry,
which is slightly higher than the empirical value -5.80677 Ry. For comparison, the wave function in which only one electron is bound whilst the other is free has the bound state energy -4 Ry  (He$^{1+}$); the difference is the ionization potential $I_0^2$ for He.

As the density of free electrons increases, screening and Pauli blocking lead to a shift of the bound state energy, and the two-electron bound state merges into the continuum at $n_e=0.56\,a_B^{-3}$.
We compare this with another test wavefunction in which one electron is bound with the wavefunction $a_1$ ($a_1=a_B/2$ at zero density), 
whilst the other electron is free ($a_2=\infty$). The bound electron experiences the charge $Z=2$ without screening by the second electron.
This state transitions into the continuum at a higher value $n_e=0.71\,a_B^{-3}$.
As can be seen from Fig. \ref{fig:ICnu}, the Mott density increases with increasing $Z$.

The symmetric solution, in which both electrons occupy the same orbit, is no longer the lowest-energy state as the density of free electrons increases. 
Due to the in-medium shifts, above $n_e^{\rm free}=0.3 \,a_B^{-3}$, the asymmetric wave function, in which one electron is found at the edge of the continuum, becomes more favorable.
If Pauli blocking is taken into account, the Pauli shift becomes smaller, as the wave function is more concentrated in coordinate space and more extended in momentum space; 
there is less overlap with the Fermi sphere. 

A preliminary discussion of the Pauli shift for the neutral He atom in a degenerate plasma was recently
presented in \cite{Ebeling26}, in which only the symmetric solution was considered. 
The approach outlined in this paper describes the dissolution of the neutral He atom in the degenerate plasma at increasing density as a stepwise ionisation process via the formation of the He$^{1+}$ ion.
We conclude that the disappearance of the bound state is not the simultaneous ionisation of both electrons, but a step-by-step, sequential ionisation as density increases.

A similar approach can be used to describe the shift of He-like ions in other materials with a charge number $Z$ of the nuclei in the medium. 
Assuming the symmetric product approach (\ref{phiHe}), the energy as a function of the parameter $a$ is given by the following (in Ry units):
\begin{equation}
    E(a)=\frac{2}{a^2}-\frac{4 Z}{a}+\frac{5}{4 a},\qquad Z^{\rm eff}=\frac{1}{a^{\rm eff}}=Z-\frac{5}{16}
\end{equation}
is obtained as the minimum of $E(a)$. For carbon, for example, the energy minimum yields
$Z^{\rm eff}=5.6875$.
As in the case of He, the pressure ionisation of C$^{4+}$ in the degenerate plasma occurs step by step, one after the other, via the formation of C$^{5+}$, due to Pauli exclusion principle.

A similar approach can be used to describe the in-medium shift of He-like ions of other materials, with charge number $Z$ of the nuclei. 
Assuming the symmetric product ansatz (\ref{phiHe}), the energy as function of the parameter $a$ follows as (Ry units)
\begin{equation}
    E(a)=\frac{2}{a^2}-\frac{4 Z}{a}+\frac{5}{4 a},\qquad Z^{\rm eff}=\frac{1}{a^{\rm eff}}=Z-\frac{5}{16}
\end{equation}
follows as the minimum of $E(a)$. For instance,  for Carbon we find from the minimum of energy.
$Z^{\rm eff}=5.6875$.
As in the case of He, the pressure ionisation of C$^{4+}$ in the degenerate plasma due to Pauli blocking occurs stepwise, subsequentally, via the formation of C$^{5+}$.

Ions containing more than two electrons can be treated in a similar way. Using a product ansatz for the many-electron wavefunction, we can introduce an effective mean-field potential. 
The resulting single-electron orbits describe the wavefunctions of the quasi-particle states, which are partially blocked due to the Pauli principle as the density of free electrons increases.
This approach to the many-electron wavefunction, which neglects the formation of correlations, reduces the calculation of energy shifts to the single-electron problem, which evolves in an effective self-consistent potential.
A simple case is that of Li-like ions, in which a third electron moves on top of the core of the $1s$ electrons, for example C$^{3+}$.
The most important prerequisite for calculating Pauli blocking is knowledge of the many-electron wavefunction of the ion; however, in this work, we cannot enter into this broad field of research, which is dominated by density functional theory (DFT).

\section{Comparison: DFT, Averaged atom model}
\label{sec:6}

We show that theories for classical plasmas lose their validity as soon as the plasma becomes degenerate.
Simulation programs such as OPAL \cite{OPAL} do not account for Pauli blocking, but this is an important effect that destroys bound states at high densities.
Experiments at high densities suggest higher ionization degrees, as predicted by standard approaches that do not account for the Pauli blocking effect \cite{Fletcher14,Kraus16,preston13}.

DFT offers an approach based on a rigorous quantum mechanical framework that takes the Pauli principle into account.
Calculations were carried out \cite{Bethkenhagen20}, but complete ionisation could not be achieved.
It remains an open question whether DFT calculations can describe the composition of a partially ionised plasma.
DFT is a quasi-particle approach in which correlations between electrons are not adequately taken into account.
As is known from the treatment of the H$_2$ molecule, in an approach where electrons move in the effective mean field of the other particles, polar states can arise that are suppressed by Coulomb repulsion.
The occurrence of bound states means that an electron is located at a specific ion and is not distributed throughout the entire system.
Such correlations may not be correctly described within the DFT formalism.

One approach that models correlation effects in a semi-empirical manner is the average-atom model; see \cite{Potekhin}.
The introduction of the Wigner–Seitz cell to solve the Schr\"odinger equation reflects the Pauli principle empirically to a certain extent, as the presence of other electrons reduces the probability that the bound electron will be found outside the Wigner–Seitz cell.
Although this model yields plausible results, it does not represent a fundamental approach.
The Pauli principle is not formulated in position space, but in the space of quantum states, such as momentum states.
Our quantum statistical theory, which uses the Green’s function method, is suitable for a more fundamental approach.

\section{Conclusions}
\label{sec:7}

We present a detailed study of Pauli blocking in degenerate plasmas that goes beyond perturbation theory.
The numerical solution of the in-medium Schr\"odinger equation yields, for a given temperature, a Mott density that is greater than the value derived from perturbation theory.
The disappearance of the bound state remains abruptly.
Values for the Mott density of the H-like bound state are given for various plasmas.

We have extended our investigation of Pauli blocking in degenerate plasmas to the case of multi-electron ions.
In particular, we have considered He and He-like ions. We have employed a shell model approach, in which a product state is assumed to consist of individual electron orbits moving within an effective mean-field potential.
The Planck–Larkin partition function is generalized to multi-electron ions.

We show that the Mott transition of the He-like bound state is not associated with the simultaneous dissociation of the two equivalent electrons as the density increases.
It is a stepwise process leading from the two-electron ion via the one-electron ion to the fully ionised ion.
The symmetric solution, in which both electrons bound to the ion are equivalent, breaks down into a solution in which one electron is free. 
The remaining electron is more strongly bound, as there is no screening by the second electron. 
An abrupt transition from the symmetric to the asymmetric solution is demonstrated.

\appendix

\section{Carbon}
\label{sec:Carbon}

We give detailed information about the calculations performed in \cite{IPD19}. 
Rydberg units: energies in Ry, lengths in $a_{\rm Bohr}=5.2918 \times 10^{-9}$ cm.
Consider Carbon with nuclear charge (proton number) $Z_{\rm C}=6$.
\begin{equation}
    T=T_{\rm Ry}=\frac{k_BT}{13.6057\,{\rm eV}},
\end{equation}
we consider WDM at $k_BT=100$ eV, $T_{\rm Ry}=7.34$. 
Particle (electrons, ions) number densities are given in atomic units
\begin{equation}
    n_{{\rm particle}, {\rm Bohr}} = n_{\rm particle} (5.2918 \times 10^{-9})^3  {\rm cm}^{3}.
\end{equation}
In particular, the plasma consists of free electrons, density $n_e$, and ions with different ionization states $Z_i$, densities $n_i$.
These ions include the neutral atom, $Z_0=0$, and the fully ionized ion $Z_{\rm C}$ what is the element order number (proton number in the nucleus).
Due to charge neutrality, we have 
\begin{equation}
    n_e= \sum_{i=0}^{Z_{\rm C}} Z_i n_i,
\end{equation}
where $Z_i=i=0, 1, \dots Z_{\rm C}$ denote the state of ionization.
The total number density of all nuclei is 
\begin{equation}
    n_{\rm nucl}^{\rm tot}=\sum_{i=0}^{Z_{\rm C}} n_i.
\end{equation}
The ionization degree 
\begin{equation}
    \bar Z= n_e/ n_{\rm nucl}^{\rm tot}
\end{equation}
gives the average number of free electrons per nuclei of the substance, here carbon C.
For simplicity, we drop in the following the index Ry, Bohr, and take all quantities in Ry units.

The most important in-medium effect at low densities is Debye screening. 
A charged particle in a plasma forms a polarization cloud, described by the polarization function.
Whereas in the parameter range considered here, the ions can be described by classical (Boltzmann) statistics, the electrons must be treated  by quantum (Fermi) statistics, so that for the screening parameter $\kappa$ in the Coulomb interaction follows
\begin{eqnarray}
    &&\kappa_{\rm ion}^2= \frac{8 \pi}{T} \sum_i^{Z_{\rm C}} Z_i^2 n_i\approx \frac{8 \pi}{T} \bar Z^2 n_a ,\nonumber \\
    &&\kappa_e^2=\frac{16 \pi^{1/2}}{T}\left(\frac{4 \pi}{T}\right)^{-3/2} \int_0^\infty\frac{dx}{x^{1/2} (e^{x-\mu^0_e/T}+1)},\nonumber \\
    &&\kappa = [\kappa_e^2(\mu, T) + \kappa_{\rm ion}^2(\bar Z, T, n_a)]^{1/2}.
\end{eqnarray}
The electron chemical potential $\mu_e$ is determined by the free electron density $n_e$, 
\begin{equation}
    n_e=\frac{1}{\pi^2}\int_0^\infty \frac{x^2 dx}{e^{(x^2-\mu^0_e)/T}+1} .
\end{equation}

The Debye screening has been improved by different works to describe correlation effects also at higher densities.
A widely used approximation was given by Stewart and Pyatt \cite{SP66}, see also \cite{EK63,Lin17}, and further references given there.
Within the Stewart-Pyatt approximation, the ion sphere radius around the ion $i$, charge $Z_i e$, is introduced,
\begin{equation}
    r_i=\left(\frac{3 Z_i}{4 \pi n_e}\right)^{1/3}. 
\end{equation}
For the shift of the energy of a charged particle in the plasma due to screening, the following expression has been proposed,
\begin{equation}
    \Delta^{\rm SP}_i=\frac{3}{2}(Z_i+1) \frac{2}{r_i}\left[\left(1+\frac{1}{(\kappa r_i)^3}\right)^{2/3}-\frac{1}{(\kappa r_i)^2}\right].
\end{equation}

In addition to this screening effect, the electrons in a degenerate plasma have an additional shift of the energy because of the formation of the Fermi hole around the electron, as a consequence of the Pauli principle. 
This effect follows from a quantum statistical approach, see, for instance, \cite{KKER,R13}, as the Hartree-Fock shift of the free electrons. 
The Hartree term is zero owing to charge neutrality, the Fock term at the continuum edge ($k=0$) is
\begin{eqnarray}
   &&\Delta^{e,\rm Fock}=\frac{4}{\pi}\int_0^\infty \frac{dp}{1+e^{p^2/T-\mu^0_e/T}}\nonumber \\
   &&=2 (T/\pi)^{1/2} {\rm Re}\left\{{\rm PolyLog}\left[(1/2),-e^{\mu^0_e/T}\right]\right\}
\end{eqnarray}
where $ {\rm PolyLog}\left[n,x\right]= {\rm Li}_{n}(x)$ is the polylogarithm function.
For the Fock shift of the bound states (we consider Carbon, $Z_{\rm C}=6$, and the ground state energy of the ion C$^{5+}$ has the energy -36 Ry, and the Bohr radius $a_{\rm C}= a_{\rm Bohr}/6$; the Hydrogen-like ground state has the wave function $\phi_5(p)=4(\pi/6)^{1/2}/3 (1+p^2/36)^{-2}$ in momentum space
\begin{eqnarray}
   &&\Delta^{\rm F, bound}=-\frac{8}{27 \pi^2} \int_0^\infty \frac{p^2 dp}{(1+p^2/36)^4} \int_0^\infty dq \nonumber \\
   && \times \left[2+\frac{T}{2pq} \ln \left(\frac{1+e^{(p-q)^2/T-\mu^0_e/T}}{1+e^{(p+q)^2/T-\mu^0_e/T}}\right)\right]
\end{eqnarray}
and the Pauli blocking term
\begin{eqnarray}
   &&\Delta^{\rm P, bound}=\frac{16}{3 \pi}\int_0^\infty\frac{p^2 dp}{(1+p^2/36)^3}\frac{1}{e^{p^2/T-\mu^0_e/T}+1}
\end{eqnarray}
The shifts are combined as follows:
\begin{eqnarray}
    &&\Delta_i = \Delta^{\rm SP}_i+\Delta^{\rm F, bound}+\Delta^{\rm P, bound}-\Delta^{\rm Fock}.
\end{eqnarray}
Note that these shifts are calculated in first order perturbation theory.

To evaluate the composition of the plasma, we consider only the bound states within the Beth-Uhlenbeck expression (it contains the "-1") and neglect the contribution of scattering states. The sum over all excited bound states is restricted to only the first excited state. [The Planck-Larkin expression contains a further term which can be neglected, but is of importance to obtain convergence if higher excited states are taken into account.]
We seek for the self-consistent solution with the density $n_6$ of the fully ionized ions, and
\begin{eqnarray}
    &&n_5=n_6 e^{\mu^0_e/T}\left[2\left( e^{E_{5,1}/T}-1\right) \Theta(E_{5,1})+8 \left(e^{E_{5,2}/T}-1\right) \Theta(E_{5,2})\right]
\end{eqnarray}
with $E_{5,1}=36-\Delta_5$, $E_{5,2}=36-27-\Delta_5$. The step function is $\Theta(x)=1$ for $x>0$, = 0 else. 
Only the first excited level (excitation energy 27 Ry) is taken into account. We neglect fine-structure separation. 
The degeneration factors are given in front of the contributions (in brackets).

Similarly, we have with the NIST data \cite{NIST} for the ionization potentials and for the average first excited level (neglecting fine-structure separation)
\begin{eqnarray}
    &&n_4=n_6 e^{2\mu^0_e/T+E_{5,1}/T}\left[\left(e^{E_{4,1}/T}-1\right) \Theta(E_{4,1})+16 \left(e^{E_{4,2}/T}-1\right) \Theta(E_{4,2})\right]
\end{eqnarray}
where $E_{4,1}=28.8178-\Delta_4$, $E_{4,2}=28.8178-22-\Delta_4$, and
\begin{eqnarray}
    &&n_3=n_6e^{3\mu^0_e/T+E_{5,1}/T+E_{4,1}/T}\left[2 \left(e^{E_{3,1}/T}-1\right) \Theta(E_{3,1})+6 \left(e^{E_{3,2}/T}-1\right) \Theta(E_{3,2})\right]
\end{eqnarray}
where $E_{3,1}=4.74066-\Delta_3$, $E_{3,2}=4.74066 - 0.5876-\Delta_3$.
Similar expressions are possible for $n_2, n_1, n_0$, but these are nearly zero in the plasma parameter range considered here.
Results are presented in Tab. \ref{Tab:iondegrC}.

\begin{table}[h]
\begin{center}
\caption{Free-electron density $n_e$, electron chemical potential $\mu^0_e$ [Ry], density of the fully ionized Carbon $n_6$ [$a_{\rm Bohr}^{-3}]$, 
density of all Carbon ions $n_{\rm nucl}^{\rm tot}$ [$a_{\rm Bohr}^{-3}]$, ionization degree $\bar Z$. Temperature $k_BT=100$ eV. \label{Tab:iondegrC}}
 \begin{tabular}{|c|c|c|c|c|}
\hline
$\log (n_e {\rm cm}^3)$ & $\mu^0_e$ & $n_6$ &  $n_{\rm nucl}^{\rm tot}$ & $\bar Z$ \\ 
\hline
23.&  -30.0952& 0.00063014& 0.00295601& 5.01302\\
23.2 & -26.6853& 0.00071164& 0.0048195& 4.87306\\
23.4& -23.2607& 0.00079334& 0.00782294& 4.75794\\
23.6&  -19.8128& 0.0008925& 0.0126506& 4.66306\\
23.8& -16.3281& 0.0010284& 0.0204013& 4.58315\\
24.& -12.7851& 0.0011919& 0.0329234& 4.50111\\
24.2& -9.1501& 0.0014806& 0.0528238& 4.44598\\
24.4& -5.37009& 0.002084& 0.0842188& 4.41967\\
24.5& -3.40138& 0.002643& 0.10602& 4.41997\\
24.6& -1.36226& 0.0035572& 0.133085& 4.43254\\
24.7& 0.76484& 0.0051735& 0.166403& 4.46321\\
24.75& 1.86801& 0.006483& 0.185613& 4.48931\\
24.8& 3.00171& 0.00841& 0.206534& 4.52725\\
24.85& 4.16948& 0.011432& 0.228805& 4.58509\\
24.9& 5.37522& 0.016693& 0.251365& 4.68292\\
24.95& 6.62325& 0.027742& 0.270202& 4.88782\\
25.& 7.91835& 0.04947& 0.28648& 5.17268\\
25.05& 9.2658& 0.08471& 0.315589& 5.26842\\
25.1& 10.6714& 0.24667& 0.323807& 5.76178\\
25.11& 10.9601& 0.31816& 0.31816& 6.\\
\hline
 \end{tabular}
\end{center}
\end{table}

\end{document}